\documentclass[aps,prb,twocolumn,superscriptaddress]{revtex4-2}
\usepackage{graphicx}
\usepackage{epstopdf}
\usepackage{amsmath}
\usepackage{appendix}
\usepackage{color}
\usepackage[colorlinks,linkcolor=blue,anchorcolor=blue,citecolor=blue,urlcolor=blue]{hyperref}
\usepackage{orcidlink}
\usepackage{mathtools} 
\usepackage{amssymb}
\usepackage{amsthm}

\begin{document}

\title{Two-dimensional transition metal selenides family $M$$_{2}$Se: A platform for superconductivity, band topology, and charge density waves}

\author{Shu-Xiang Qiao\orcidlink{0009-0008-7092-3333}} 
\affiliation{School of Physics and Physical Engineering, Qufu Normal University, Qufu 273165, China}

\author{Kai-Yue Jiang\orcidlink{0009-0007-0395-9662}} 
\affiliation{School of Physics and Physical Engineering, Qufu Normal University, Qufu 273165, China}

\author{Yu-Lin Han\orcidlink{0009-0001-6308-2977}} 
\affiliation{School of Physics and Physical Engineering, Qufu Normal University, Qufu 273165, China}

\author{Na Jiao} 
\affiliation{School of Physics and Physical Engineering, Qufu Normal University, Qufu 273165, China}

\author{Ying-Jie Chen} 
\affiliation{School of Physics and Physical Engineering, Qufu Normal University, Qufu 273165, China}

\author{Hong-Yan Lu\orcidlink{0000-0003-4715-7489}}\email[E-mail: ]{hylu@qfnu.edu.cn} 
\affiliation{School of Physics and Physical Engineering, Qufu Normal University, Qufu 273165, China}

\author{Ping Zhang}\email[E-mail: ]{zhang$\_$ping@iapcm.ac.cn} 
\affiliation{School of Physics and Physical Engineering, Qufu Normal University, Qufu 273165, China}
\affiliation{Institute of Applied Physics and Computational Mathematics, Beijing 100088, China}

\begin{abstract}  	
 MXenes and MBenes, which are two-dimensional (2D) transition metal carbides/nitrides and borides, have been extensively studied for their impressive properties. Recently, we reported a family of transition metal sulfides MSene ($M$$_{2}$S) with rich properties [Phys. Rev. B \textbf{111}, L041404 (2025)], it is worth studying whether selenides with similar structure also have rich properties. In this work, through high-throughput screening, we present a novel family of 2D transition metal selenides, $M$$_{2}$Se. In this family, there are fifty-eight candidate materials, of which ten are stable and metallic. Notably, eight exhibit superconductivity, among which four are superconducting topological metals. Besides, eight show charge density wave (CDW) behavior, among which five also exhibit antiferromagnetism. It is revealed that CDW originates from electron-phonon coupling rather than Fermi surface nesting. Moreover, strain can be applied to regulate the competition between CDW and superconductivity. Our findings reveal the rich properties of superconductivity, band topology, CDW, and magnetism in $M$$_{2}$Se, providing a new platform for the controllable integration of multifunctional quantum states.

\end{abstract} 

\maketitle

\section{Introduction}

Two-dimensional (2D) materials have revolutionized condensed matter physics and materials science, particularly since the successful isolation of graphene \cite{c-exp,c}. Characterized by their atomic thickness, high surface area, and tunable electronic, magnetic, and mechanical properties, these materials have significantly advanced our understanding of low-dimensional systems. Over the past two decades, numerous 2D materials have been discovered, including transition metal dichalcogenides (TMDs) \cite{tmd1,tmd2,tmd4,tmd5}, MXenes \cite{mxene1,mxene2,mxene3,mxene4}, MBenes \cite{mbene1,mbene3}, and MOenes \cite{moene1,moene2,moene3}. Each of these material classes exhibits distinct structural and functional attributes. These materials have driven significant advancements in fields such as energy storage \cite{storage1,storage2}, catalysis \cite{catalysis1,catalysis2}, optoelectronics \cite{optoelectronics1,optoelectronics2}, and quantum information science \cite{quantum}.

Among the prominent 2D materials, MXenes stand out for their high electrical conductivity, chemical tunability, and mechanical flexibility \cite{mxene1,mxene5,mxene7}. These properties enable MXenes to exhibit diverse functionalities, such as magnetism, superconductivity, and topological properties, making them a versatile platform for both fundamental studies and practical applications \cite{mxene8,c2,topology,mag1,mag2}. Similarly, the recent emergence of MBenes, the boron-based analogs of MXenes, has expanded the material design space. These 2D boron-containing materials exhibit intriguing properties, including enhanced catalytic activity and anisotropic electronic structures \cite{mbene4,mbene5}. Both materials demonstrate the potential of B, C, N to stabilize transition metal compounds. Our recent work on the MSene family has shown that S can also stabilize transition metal compounds with MXene-like structure \cite{msene}. The MSene family possesses diverse physical properties, such as magnetism, coexistence of superconductivity and topology, and competition between superconductivity and CDW \cite{qiao2025}.

TMDs, such as MoS$_{2}$ and WSe$_{2}$, have demonstrated that chalcogen elements play a crucial role in stabilizing 2D materials and enabling diverse electronic and magnetic behaviors \cite{tmd1,tmd8,tmd6,tmd7}. Driven by their strong spin-orbit coupling (SOC), layered structures, and robust electron-phonon interactions, TMDs are also known to host collective phenomena, such as superconductivity, charge density waves (CDW), and nontrivial topological states \cite{tmd9,tmd10,tmd11,mos2app2,mos2app3}. These findings raise an intriguing question: can Se atoms stabilize MXene-like structure in a manner similar to carbon or boron, while also preserving or even enhancing these rich physical properties?

\begin{figure*}
	\centering
	\includegraphics[width=17cm]{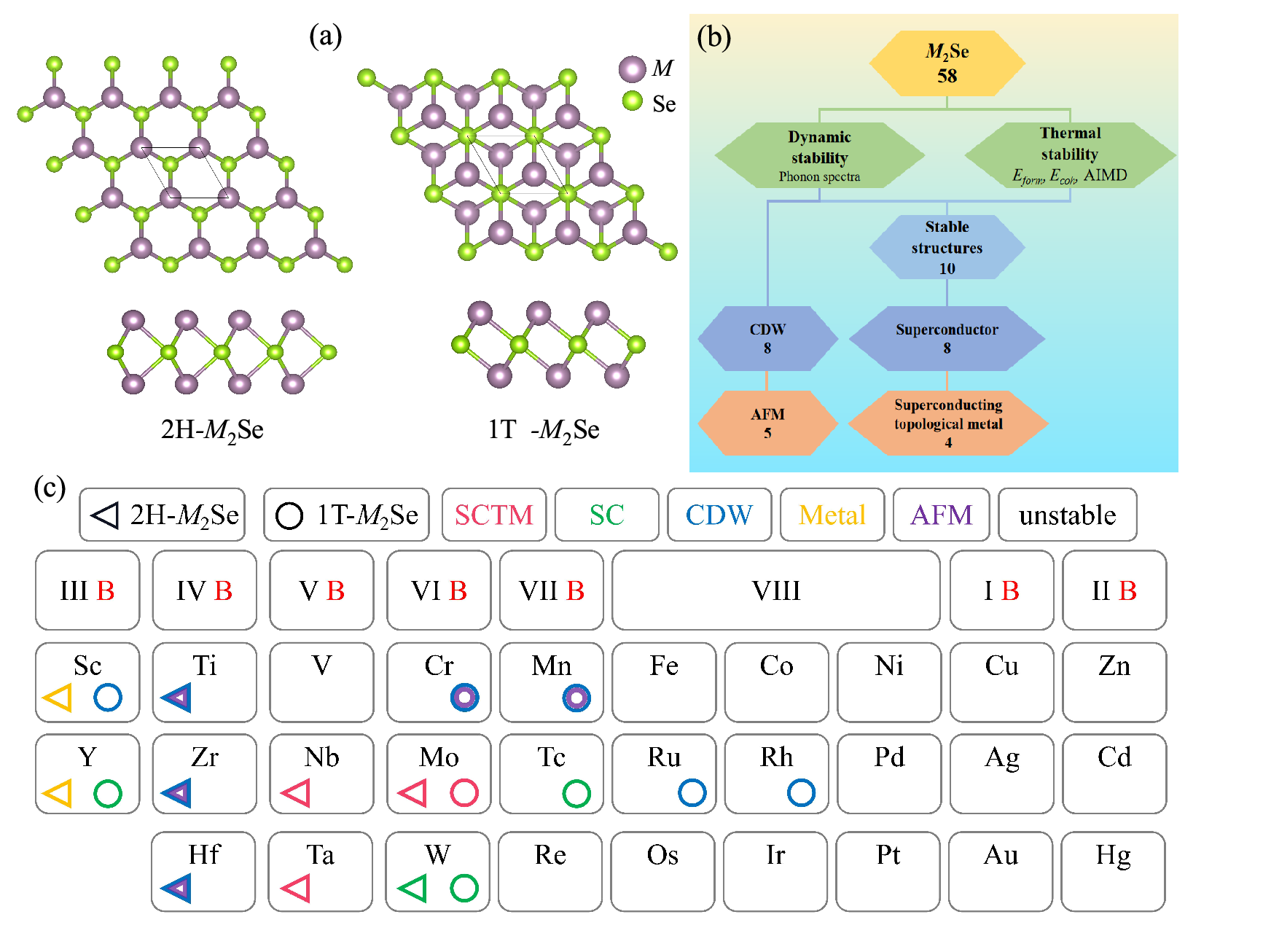}
	\caption{ (a) Top and side views of 2H-$M$$_{2}$Se (left) and 1T-$M$$_{2}$Se (right), (b) Schematic diagram of the screening procedure for $M$$_{2}$Se, (c) Distribution of stable $M$$_{2}$Se in the periodic table. For clarity, different properties are represented by different colors. In (c), Metal labelled by yellow represents the common metal without superconductivity and band topology.}
	\label{1} 
\end{figure*}

In this study, through high-throughput screening, we propose a novel family of 2D transition metal selenide $M$$_{2}$Se with MXene-like structure. Among the fifty-eight potential candidates identified, ten of which are stable in both the 2H or 1T phases. This new family displays rich properties, including superconductivity, superconducting topological metals (SCTMs), CDW and antiferromagnetic (AFM) phases. These findings establish chalcogen-stabilized MXene-like materials as a promising platform for investigating the competition and coexistence of various orders in 2D systems. By providing a systematic analysis of their structural and electronic properties, we aim to guide experimental efforts in synthesizing and characterizing these compounds, thereby advancing our understanding of the interplay between multiple quantum orders in low-dimensional systems.

\section{Computational methods}

The structural relaxation and electronic property calculations were carried out within the framework of density functional theory (DFT) using the Vienna ab-initio simulation package (VASP) \cite{1} and the software QUANTUM-ESPRESSO (QE) \cite{2}. Phonon spectrum and electron-phonon interaction were performed within density functional perturbation theory (DFPT) \cite{66,67} as implemented in QE. The exchange-correlation functional was described through generalized gradient approximation (GGA) with Perdew–Burke–Ernzerhof (PBE) parametrization \cite{3}. Electron-ion interactions were modeled using the projector augmented-wave (PAW) method \cite{4}. The edge states were determined using the iterative Green's function approach, implemented in the WANNIERTOOLS package \cite{5,6}, with the basis set derived from the maximally localized Wannier functions (MLWFs) \cite{7,8}, obtained via the VASP2WANNIER90 interface \cite{9}. Further computational details can be found in the Supplemental Material (SM) \cite{SM}.

\section{Results and discussions}
\subsection{Lattice structure and Stability}

Figure \ref{1}(a) shows the top and side views of 2H- and 1T-$M$$_{2}$Se. Both of them possess a sandwich structure, with a layer of Se atoms sandwiched between two layers of transition metal atoms, similar to MXene. For 2H-$M$$_{2}$Se, the structure adopts a triangular prism coordination with hexagonal symmetry, which belongs to the $P$-$6m2$ space group ($D$$_{3h}$ point group). The atoms stack along the $z$-axis in an ABA pattern, where A represents two layers of transition metal sites, and B represents the Se layer, as shown in Fig. \ref{1}(a). In 1T-$M$$_{2}$Se, the structure exhibits octahedral coordination and tetragonal symmetry with $P$-$3m1$ space group ($D$$_{3d}$ point group). The atoms stacking follows ABC pattern, where A and C represent two transition metal layers, and B represents the intermediate Se layer, as shown in Fig. \ref{1}(a). The primitive cells of both the 2H and 1T-$M$$_{2}$Se contain three atoms: two transition metal atoms and one Se atom.

Based on the elemental composition of TMDs and the elemental ratio of MXenes, fifty-eight $M$$_{2}$Se candidate materials are proposed for both 2H and 1T phases. Their dynamics stability is analyzed using phonon spectra, and thermal stability are analyzed though calculating formation energy $E_{form}$, cohesive energy $E_{coh}$ and $ab$ $initio$ molecular dynamics (AIMD), as shown in the flowchart in Fig. \ref{1}(b).

 Firstly, the phonon spectra of fifty-eight structures are calculated. The absence of imaginary frequencies indicates their dynamic stability, as shown in Fig. S1 of the SM \cite{SM}, which is crucial for evaluating new structures. Among them, there are eleven dynamically stable materials. Secondly, the formation energies $E_{form}$ of dynamically stable structures are defined as $E_{form}$ = [$E_{tot}$ $-$ (2$E_{M}$+$E_{Se}$)]/3 and calculated, where $E_{tot}$, $E_{M}$, and $E_{Se}$, are the total energies of $M$$_{2}$Se, the free metal element $M$, and body centered cubic Se which is experimentally synthesised, respectively. Among the eleven dynamically stable materials, ten have negative formation energies, confirming their thermodynamic stability, which listed in Table S1 \cite{SM}. Additionally, the cohesive energies $E_{coh}$ are also calculated and listed in Table S2 and Fig. S2 of SM \cite{SM}. These values are lower than some experimentally prepared materials \cite{cu2si1,cu2si2,cu2ge,gas}, which also indicates their thermal and bond stability. Finally, the AIMD at 300 K are performed for the structures with stable phonon spectra and cohesive energies, as presented in Fig. S3 \cite{SM}. The structures remain stable and do not deform, indicating their thermodynamic stability. In summary, after above high-throughput screening, a total of ten stable materials and eight CDW materials are identified, as shown in Fig. \ref{1}(b).

\begin{figure*}
	\centering
	\includegraphics[width=16cm]{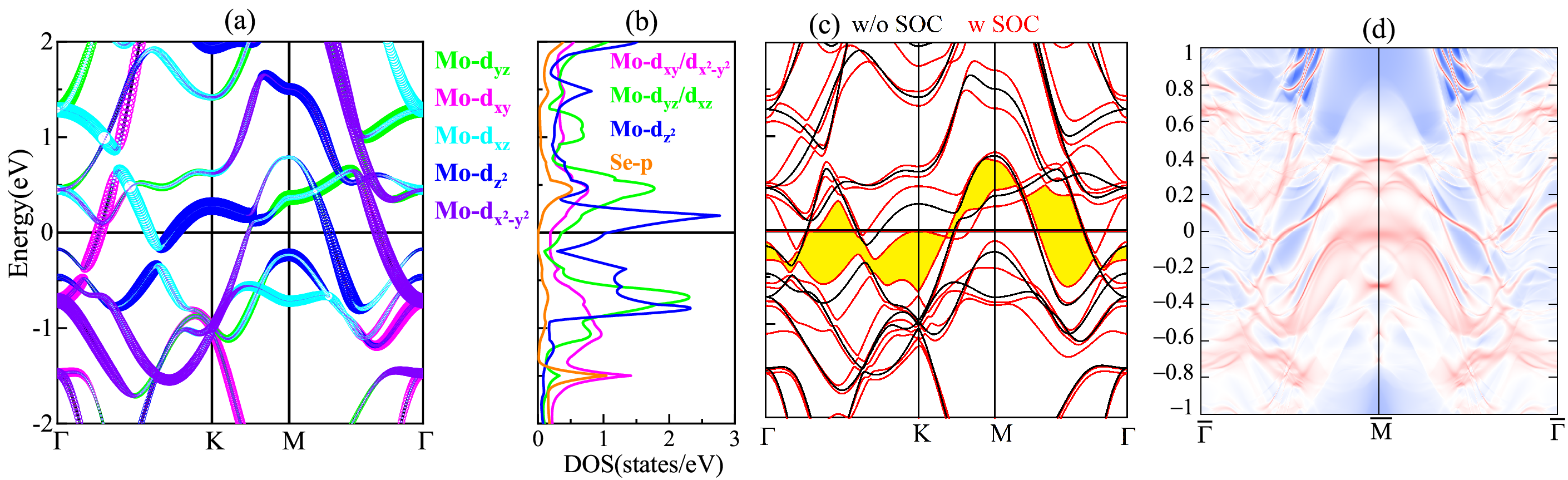}
	\includegraphics[width=16cm]{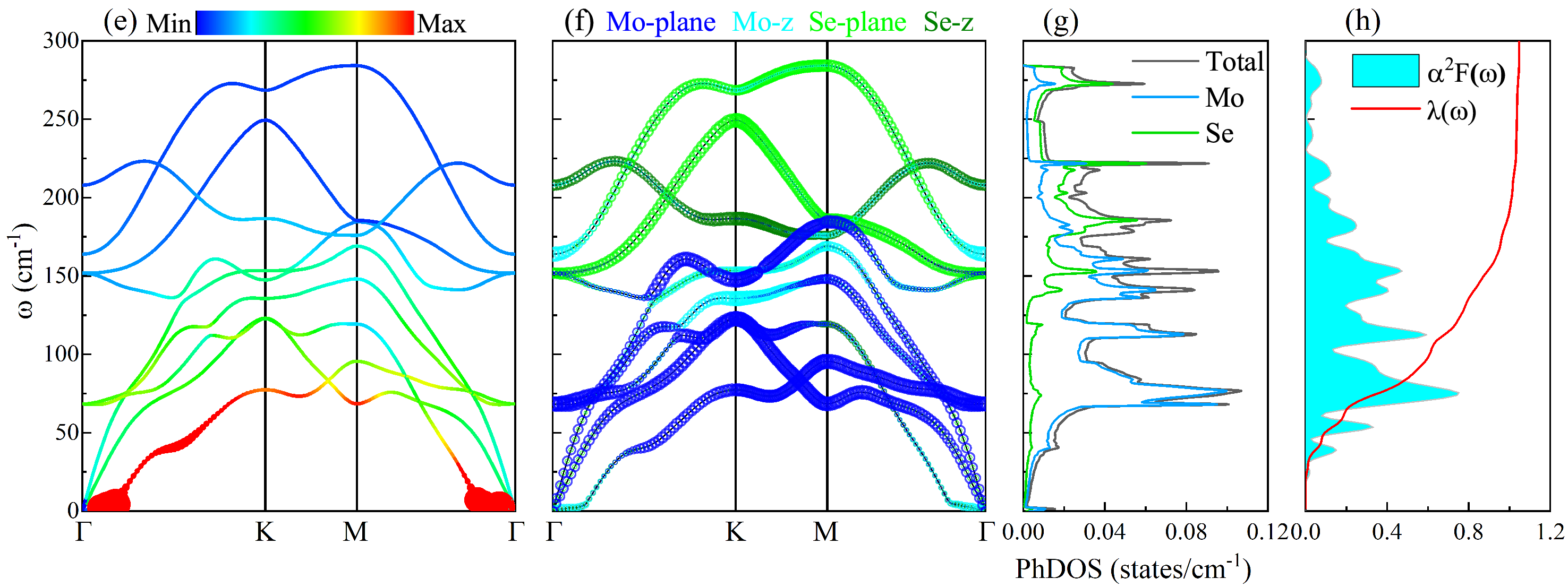}
	\caption{ Band topology and superconducting properties of 2H-Mo$_{2}$Se. (a) Orbital-projected electronic band structure without SOC along high-symmetry line $\Gamma$-$K$-$M$-$\Gamma$. (b) PDOS without SOC. (c) Band without (black) and with (red) SOC. (d) Edge states. (e) Phonon dispersion weighted by the magnitude of $\lambda_{\textbf{q}\nu}$ (EPC for phonon mode ${\textbf{q}\nu}$). (f) Phonon dispersion weighted by the vibration modes of each atom. (g) Phonon DOS. (h) Eliashberg spectral function $\alpha^{2}F (\omega)$ and EPC $\lambda (\omega)$.}
	\label{2}
\end{figure*}

Next, the structural parameters, electronic properties, superconducting properties of the stable materials are analyzed. Their structural parameters and bonding properties are summarized in Table S3 and Fig. S4 of the SM \cite{SM}, including lattice parameters $a$, $M$-Se bond length, interlayer distance $h$, charge density difference, Bader charge, and electron localization function (ELF). These results show that the transition metal atoms lose electrons while Se atoms gain electrons. The bonds between the $M$ and Se atoms exhibit ionic bonding characteristics, because the charge density around Se atoms is greater than 0.8, and is higher than that around $M$ atoms, as reported for Ti$_{2}$BN \cite{ti2bn}. Energy band and density of states (DOS) calculations for the ten materials, shown in Fig. S5 of the SM \cite{SM}, reveal that all stable materials exhibit metallic states, allowing for the study of their electronic topology and superconducting properties.

\begin{table*}
	\caption{The logarithmically averaged phonon frequency $\omega_{log}$, total EPC constant $\lambda$, $T_{c}$, Sommerfeld constant $\gamma$, specific heat jump at $T_{c}$ ${\Delta C\left(T_c\right)}$, superconducting gap at 0 K ${\Delta(0)}$, critical magnetic field at 0 K $H_c(0)$, isotope coefficient $\alpha$, and $\frac{2 \Delta(0)}{k_{{B}} T_{{c}}}$ for superconducting $M$$_{2}$Se.}
	\centering
	\setlength{\tabcolsep}{4pt}
	\renewcommand{\arraystretch}{1}
	\begin{tabular}{ccccccccccccccccc}
		\hline
		\hline
		Materials   & $\omega_{log}$ (K) & $\lambda$ & $T_{c}$ (K) & $\gamma$ ($\mathrm{\frac{mJ}{mol \cdot K^{2}}}$) & ${\Delta C\left(T_c\right)}$ ($\mathrm{\frac{mJ}{mol \cdot K^{2}}}$) & ${\Delta(0)}$ (meV) & $H_c(0)$ (T)  & $\alpha$ & $\frac{2 \Delta(0)}{k_{{B}} T_{{c}}}$ \\ \hline
		2H-Nb$_{2}$Se &134.71	&0.56	&2.4	    &0.111	&0.167	&0.376	&0.020	&0.44	&3.58 \\ 
		2H-Mo$_{2}$Se &131.51	&1.05	&10.4	    &0.173	&0.367	&1.809	&0.112	&0.48	&4.04 \\ 
		2H-Ta$_{2}$Se &122.72	&0.37	&0.4	    &0.101	&0.145	&0.061	&0.003	&0.36	&3.53 \\ 
		2H-W$_{2}$Se  &166.88	&0.47	&1.6	    &0.109	&0.158	&0.249	&0.013	&0.42	&3.55 \\ 
		1T-Y$_{2}$Se  &114.83	&0.36	&0.4	    &0.091	&0.131	&0.061	&0.003	&0.35	&3.53 \\ 
		1T-Mo$_{2}$Se &168.09	&0.71	&6.1	    &0.131	&0.216	&0.967	&0.055	&0.46	&3.68 \\ 
		1T-Tc$_{2}$Se &177.17	&0.62	&4.5	    &0.124	&0.193	&0.702	&0.039	&0.45	&3.62 \\ 
		1T-W$_{2}$Se  &170.28	&0.55	&2.9	    &0.097	&0.144	&0.442	&0.022	&0.44	&3.57 \\ 
		\hline \hline
	\end{tabular}
	\label{sc}
\end{table*}

After considering the electronic and electron-phonon coupling (EPC) calculation, the diverse properties of $M$$_{2}$Se are summarized in Fig. \ref{1}(c). There are ten stable materials and eight CDW materials. Among the ten stable materials, four are superconductors, four are SCTMs, and two are metallic. Five of eight CDW materials exhibit AFM properties. Due to the large number of materials, we focus on representative ones with superior properties (e.g., high $T_c$ and rich quantum orders). Specifically, we analyze 2H-Mo$_2$Se as a SCTM and 1T-Ru$_2$Se/Sc$_2$Se as CDW materials to study their CDW origin and relation to superconductivity. The properties of the remaining materials are provided in the SM \cite{SM}.

\subsection{Band topology and superconductivity}

Recently, SCTMs have attracted significant research interest due to their unique topological boundary states and the potential applications of quantum fault-tolerant computers. The band topology and superconducting properties of 2H-Mo$_{2}$Se are shown in Fig. \ref{2}. As shown in Figs. \ref{2}(a) and \ref{2}(b), the orbital-projected electronic band without SOC and the partial DOS (PDOS) of 2H-Mo$_{2}$Se indicate that the electronic states at the Fermi level are primarily contributed by the Mo-$d_{z^2}$ orbitals, followed by the Mo-$d_{yz}$ and Mo-$d_{xz}$ orbitals, and the contribution of Se is minimal, which is a characteristic of transition metal compounds. Interestingly, a close point can be found between $\Gamma$ and $M$, which is a typical feature of topological materials, as reported in TiB$_{4}$ \cite{tib4} and W$_{2}$N$_{3}$ \cite{w2n3}. As shown in Fig. \ref{2}(c), the close point opens up with a 10 meV energy gap with the effect of SOC, forming a continuous gap throughout the Brillouin zone (highlighted in yellow in Fig. \ref{2}(c)), which is essential for establishing the topological invariant $\mathbb{Z}_{2}$. The calculated topological invariant $\mathbb{Z}_{2}$ is 1, indicating that 2H-Mo$_{2}$Se is topologically non-trivial. Moreover, the existence of topological edge states further confirms that 2H-Mo$_{2}$Se is a topological material, as shown in Fig. \ref{2}(d).

\begin{figure}
	\centering
	\includegraphics[width=8cm]{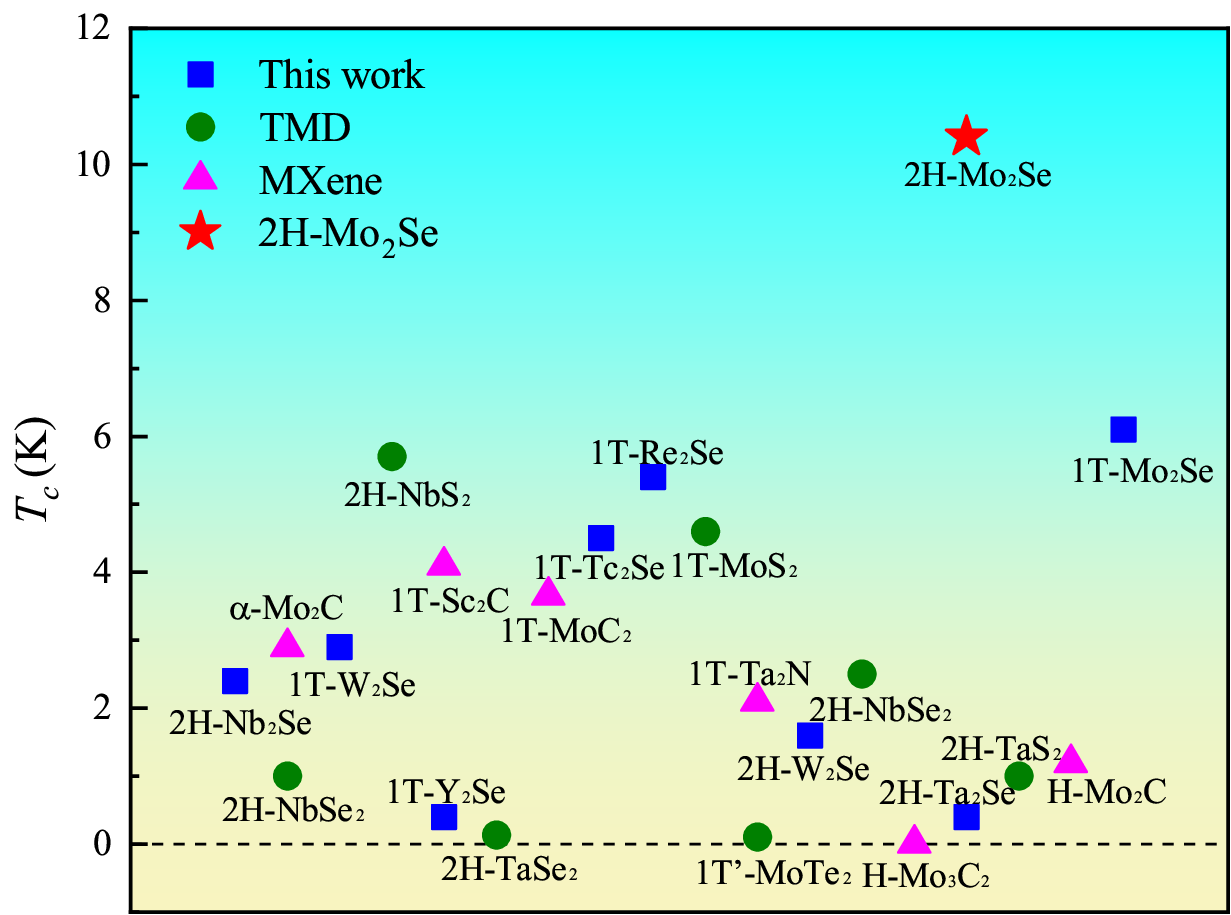}
	\caption{ $T_{c}$ of TMDs and MXenes superconductors and the $M$$_{2}$Se superconductors predicted in this work.}
	\label{3}
\end{figure}

\begin{figure*}
	\centering
	\includegraphics[width=17cm]{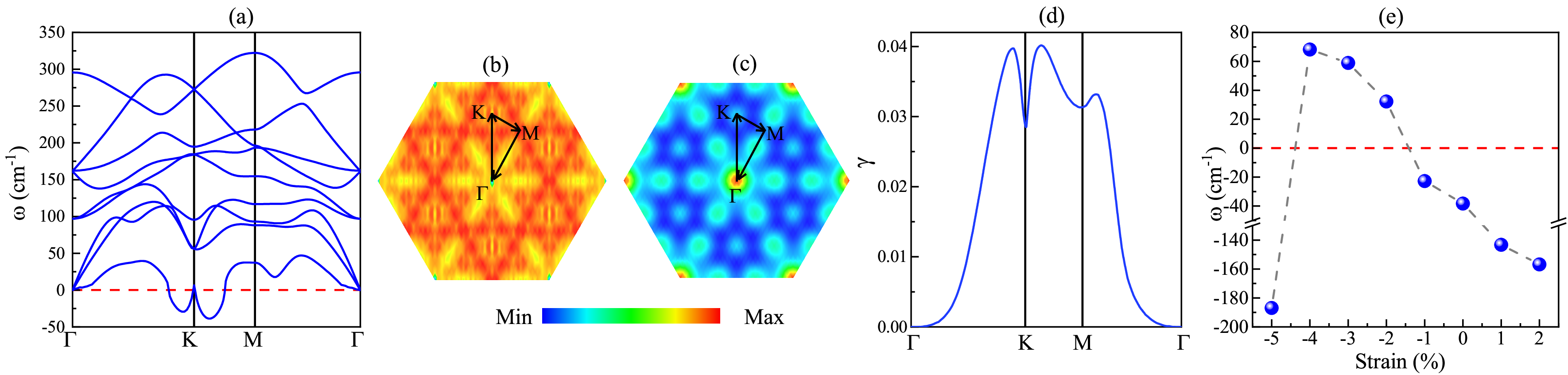}
	\includegraphics[width=17cm]{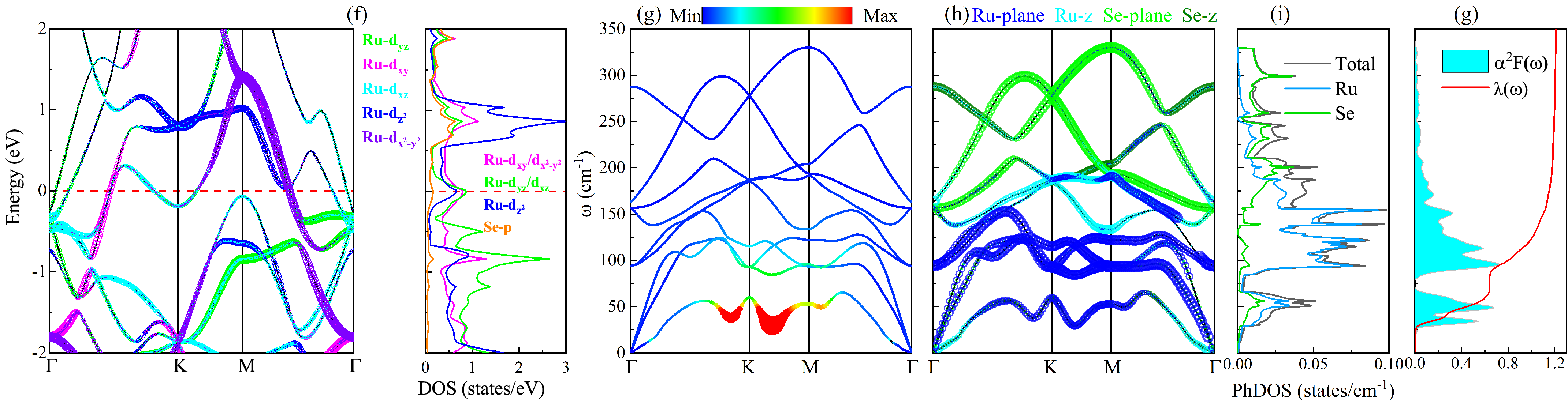}
	\caption{ (a) Phonon spectrum, (b) real part and (c) imaginary part of the electronic susceptibility, and (d) phonon linewidth of 1T-Ru$_{2}$Se. (e) Changes of the lowest mode frequency with strain. (f) Orbital-projected band structure and PDOS. (g) Phonon dispersion weighted by the magnitude of $\lambda_{\textbf{q}\nu}$ (EPC for phonon mode ${\textbf{q}\nu}$). (h) Phonon dispersion weighted by the vibration modes of each atom. (i) Phonon DOS. (g) Eliashberg spectral function $\alpha^{2}F (\omega)$ and EPC $\lambda (\omega)$ of 2$\%$ compressive strained 1T-Ru$_{2}$Se.}
	\label{4}
\end{figure*}

Since 2H-Mo$_{2}$Se is metallic, its possible phonon-mediated superconducting properties have also been investigated. The region with the strongest EPC lies below 75 cm$^{-1}$, primarily due to the in-plane vibrations of Mo atoms. Due to the larger atomic mass of Mo compared with Se, the vibrations of Mo atoms occur mainly in the low-frequency range ($w$ $<$ 175 cm$^{-1}$), while Se atoms primarily vibrate in the high-frequency range ($w$ $>$ 175 cm$^{-1}$), as shown in Figs. \ref{2}(f) and \ref{2}(g). Combining Figs. \ref{2}(e), \ref{2}(f) and \ref{2}(h), the strong EPC from the in-plane vibration of Mo atoms accounts for approximately 68.2$\%$ to the total EPC constant $\lambda$ in the low-frequency (25 cm$^{-1}$ $<$ $w$ $<$ 75 cm$^{-1}$). Then, the calculated EPC constant $\lambda$ and $T_{c}$ of 2H-Mo$_{2}$Se are 1.05 and 10.4 K, respectively. Its superconductivity mainly originates from the coupling between Mo-$d_{z^2}$ electrons and the low-frequency in-plane vibration of Mo atoms, similar to that in other transition metal superconductors \cite{jiaona}. As mentioned above, 2H-Mo$_{2}$Se is a promising candidate for superconducting topological materials \cite{acpd3,jiang}. Its topological properties provide strong resilience to interference, low energy consumption, and high-speed operation, making it a promising candidate for quantum computing and communication applications. Moreover, possible topological superconductivity can be further explored in it.

For the other SCTMs, their band structures without and with SOC as well as their edge states are presented in Fig. S6 of the SM \cite{SM}. The superconducting properties of the remaining superconductors are presented in Fig. S7 \cite{SM}, including the phonon spectra weighted by the magnitude of $\lambda_{\textbf{q}\nu}$, the vibration modes of each atom, phonon DOS and Eliashberg spectral function $\alpha^{2}F (\omega)$ and EPC $\lambda (\omega)$. The superconducting properties of these materials are summarized in Table \ref{sc}, including logarithmically averaged phonon frequency $\omega_{log}$, total EPC constant $\lambda$, superconducting critical temperature $T_{c}$, sommerfeld constant $\gamma$, specific heat jump at $T_{c}$ ${\Delta C\left(T_c\right)}$, superconducting gap at 0 K ${\Delta(0)}$, critical magnetic field at 0 K $H_c(0)$, isotope coefficient $\alpha$, and $\frac{2 \Delta(0)}{k_{{B}}T_{{c}}}$. The calculated $\frac{2\Delta(0)}{k_{{B}}T_{{c}}}$ and $\alpha$ of $M$$_{2}$Se closely match those predicted by conventional BCS theory ($\frac{2\Delta(0)}{k_{B} T_{c}} \approx 3.53$ and $\alpha \approx 0.5$), confirming that these $M$$_{2}$Se are conventional phonon-mediated superconductors. Moreover, the $T_{c}$s of several $M$$_{2}$Se are higher than that of most TMDs \cite{sm16,sm17,sm18,sm19,sm20,sm21,sm22,sm23,sm24} and MXenes \cite{sm25,sm26}, as shown in Fig. \ref{3} and Table S4 \cite{SM}. Notably, the $T_{c}$ of 2H-Mo$_{2}$Se exceeds 10 K, which is rarely achieved for intrinsic TMDs and MXenes.

\subsection{CDW property}

CDW is an interesting quantum phenomena where solids undergo periodic lattice distortions at low temperatures, accompanied by periodic modulation of charge density. As shown in Fig. \ref{4}(a), the primitive cell of 1T-Ru$_{2}$Se contains three atoms, which give rise to nine phonon modes with three acoustic modes and six optical modes. The appearance of imaginary frequencies near $K$ implies the dynamic instability of 1T-Ru$_{2}$Se. Notably, the positions of these imaginary frequencies, known as Kohn anomalies, occur at special points, which is a typical characteristic of CDW materials \cite{mosh,tate2}. Thus, 1T-Ru$_{2}$Se possesses CDW properties. In recent years, CDWs in 2D monolayers have attracted significant attention worldwide, spurred by the growing interest in 2D materials \cite{tate2,luwenjian,cdw2,cdw3,cdw4}.

Currently, several mechanisms have been proposed to explain the origin of CDWs in 2D transition metal compounds. Many studies suggest that CDW may originate from Fermi surface nesting (FSN) or EPC \cite{mosh,tas2,tasi2n4,tate2}. The intensity of FSN is assessed by calculating the real and imaginary parts of the electronic susceptibility, while the EPC strength is quantified by calculating the phonon linewidth. The real part of electronic susceptibility reflects the stability of the electronic system, while the imaginary part is associated with the electronic excitations and topological properties \cite{2008}. If CDW originates from the FSN, the peaks of the imaginary part must coincide with those of the real part and the CDW wave vectors \cite{2006}. It is found that the CDW wave vector is located around the $K$ point, the peak of the real part is near the $M$ point, and the peak of the imaginary part lies between $\Gamma$ and $M$, as shown in Figs. \ref{4}(a), \ref{4}(b) and \ref{4}(c). Since these peaks do not align to each other, the CDW in 1T-Ru$_{2}$Se does not originate from FSN. Additionally, Kohn anomalies may also originate from EPC, with their strength described by the phonon linewidth, as illustrated in Fig. \ref{4}(d). It is obvious that the peak of phonon linewidth corresponds precisely to the CDW wave vectors, confirming that the CDW in 1T-Ru$_{2}$Se originates from EPC \cite{cdw3,cdw4}. Besides 1T-Ru$_{2}$Se, the origins of the other non-magnetic CDW materials (1T-Sc$_{2}$Se and 1T-Rh$_{2}$Se) are also attributed to EPC, as shown in Fig. S8 of the SM \cite{SM}.

Above study shows that the CDW of 1T-Ru$_{2}$Se is driven by EPC, with similar mechanism in traditional BCS superconductors. Thus, there may be a competition between CDW and superconductivity in 1T-Ru$_{2}$Se. Previous studies have shown that biaxial strain can be applied to 2D CDW materials to suppress CDW and generate superconductivity \cite{mosh,tas2,tasi2n4,cdw3,cdw4}. Moreover, certain levels of strain are often introduced during the experimental preparation process, thus, we investigate the effects of biaxial strain on 1T-Ru$_{2}$Se. Figure \ref{4}(e) shows the changes in the lowest mode frequency under varying biaxial strain, and the phonon spectra under strains are shown in Fig. S10 \cite{SM}. The results indicate that a compressive strain of 2$\%$ to 4$\%$ effectively suppresses the CDW phase and stabilizes the structure.

For 2$\%$ to 4$\%$ compressive strained 1T-Ru$_{2}$Se, their superconductivity are investigated, here, superconductivity under 2$\%$ compressive strain is analyzed in detail. Its electronic states near the Fermi level are mainly derived from the Ru-$d_{yz}$/$d_{xz}$ orbitals, followed by the Ru-$d_{xy}$/$d_{x^2-y^2}$ orbitals, as presented in Fig. \ref{4}(f). Figures \ref{4}(g) to \ref{4}(j) illustrate the superconductivity of 2$\%$ compressive strained 1T-Ru$_{2}$Se. Strong EPC is observed at the Kohn anomalies near the $K$ with the range from 25 to 50 cm$^{-1}$, which originates from the in-plane vibrations of Ru. As shown in Figs. \ref{4}(h) and \ref{4}(i), the vibrations of Ru atoms are predominantly in the low-frequency region, while the vibration of Se atoms is primarily in the high-frequency range, which is due to the larger mass of Ru compared to Se. Furthermore, the EPC arising from in-plane vibrations of Ru in the low-frequency region ($w$ $<$ 75 cm$^{-1}$) contributes the approximately 62.5$\%$ of the EPC constant $\lambda$, while the vibrations in the 75-125 cm$^{-1}$ range contribute an additional 34.9$\%$. From the above analysis, the superconductivity of 2$\%$ biaxial compressive strained 1T-Ru$_{2}$Se mainly results from the coupling between Ru-$d_{yz}$/$d_{xz}$ electrons and the low-frequency in-plane vibrations of Ru. The total EPC constant $\lambda$ and $T_{c}$ of 2$\%$ biaxial compressive strained 1T-Ru$_{2}$Se are 1.19 and 9.1 K, respectively.

\begin{figure}
	\centering
	\includegraphics[width=8cm]{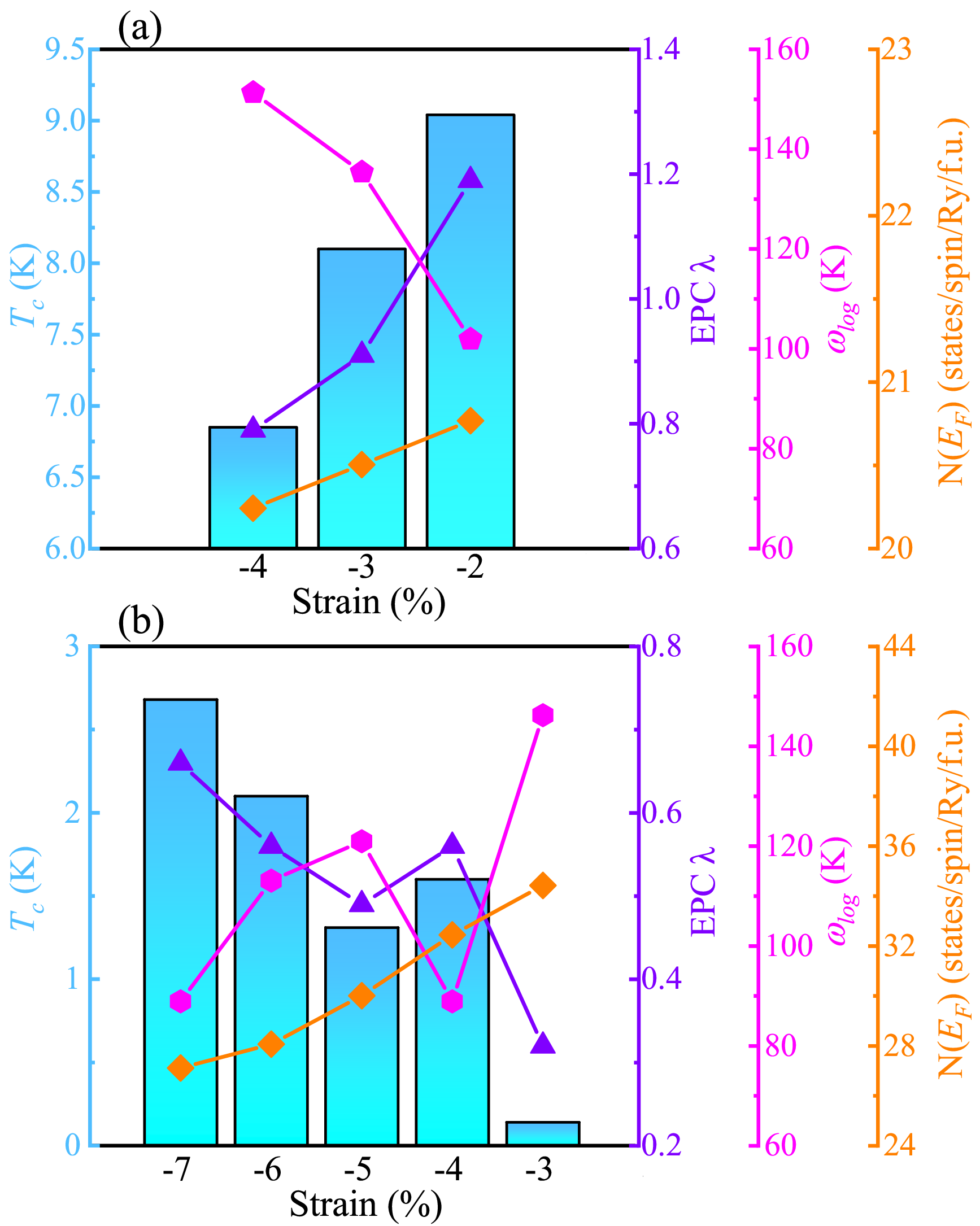}
	\caption{ Variation of $T_{c}$, EPC $\lambda$, $\omega_{log}$, and $N(E_{F})$ under strains for (a) 1T-Ru$_{2}$Se and (b) 1T-Sc$_{2}$Se.}
	\label{5}
\end{figure}

Additionally, the phonon and superconducting properties of CDW materials under different biaxial strains are summarized in Fig. \ref{5} and Fig. S9 \cite{SM}. Compressive strain is found to suppress the CDW in 1T-Ru$_{2}$Se and 1T-Sc$_{2}$Se, thereby inducing superconductivity. However, the CDW in 1T-Rh$_{2}$Se remains stable under biaxial strain, and the biaxial strain cannot suppress its CDW, its phonon spectra under biaxial compressive strain are shown in Fig. S12 \cite{SM}. For 1T-Ru$_{2}$Se, the CDW is suppressed under 2$\%$ to 4$\%$ biaxial compressive strain, and its superconductivity exhibits a monotonic trend. The higtest $T_{c}$ of 9.1 K is achieved at 2$\%$ biaxial compressive strain, as shown in Fig. \ref{5}(a). For 1T-Sc$_{2}$Se, its CDW is suppressed under biaxial compressive strains ranging from 3$\%$ to 7$\%$, which is a broader range than that for 1T-Ru$_{2}$Se, as shown in Figs. S9 and S11 \cite{SM}. Notably, the superconductivity in biaxial compressively strained 1T-Sc$_{2}$Se follows two-dome like behavior \cite{dome1,dome2,dome3}, as illustrated in Fig. \ref{5}(b), with the highest $T_{c}$ of 2.7 K at 7$\%$ compressive strain. For the other CDW materials, incliding 2H-Ti$_{2}$Se, 2H-Zr$_{2}$Se, 2H-Hf$_{2}$Se, 1T-Cr$_{2}$Se, and 1T-Mn$_{2}$Se, their magnetic ground states are determined to be A-AFM by comparing the energies of A-AFM, Col.-FM, Col.-AFM, Zig.-FM, and Zig.-AFM relative to FM structure, as shown in Fig. S13 and Table S5 \cite{SM}. Due to their magnetism, their complex interplay between magnetism, CDW, and superconductivity will be investigated in future work.

Based on the above research, it can be concluded that the origins of CDW in 1T-Rh$_{2}$Se, 1T-Ru$_{2}$Se and 1T-Sc$_{2}$Se are EPC. The CDW can be suppressed by applying appropriate biaxial compressive strain, which can also modulate the competition with superconductivity. These findings provide theoretical groundwork for further experiments.

\subsection{Discussion}

The newly predicted $M$$_{2}$Se family represents an innovative class of 2D transition metal selenides, which uniquely combines the elemental composition of TMDs with the stoichiometric ratios characteristic of MXenes. This hybrid nature endows the $M$$_{2}$Se family with several distinct advantages over both TMDs and MXenes. Compared to TMDs, this family not only retains excellent electrical conductivity but also expands the range of metallic materials available, thus offering a versatile platform for exploring a diverse array of electronic properties, including superconductivity, band topology, and magnetism. Unlike MXenes, $M$$_{2}$Se family exhibits rich CDW properties, which can give rise to the competition between CDW and superconductivity, and complex interplay between AFM and CDW. Moreover, the $M_{2}$Se family boasts higher intrinsic $T_{c}$ than either TMDs or MXenes. As a hybrid of TMD and MXene, the $M$$_{2}$Se family not only preserves and integrates their rich physical properties but also enhances these properties, surpassing the capabilities of each material individually. Therefore, our findings demonstrate that the $M$$_{2}$Se family possesses a wealth of physical properties, making it an ideal platform for the controllable integration of multifunctional quantum states.

\section{Conclusion}

In this study, we have systematically investigated a class of 2D transition metal selenides ($M$$_{2}$Se) using first-principles calculations, revealing their structural stability, topological superconductivity, and CDW properties. Through comprehensive screening of fifty-eight candidate materials, we identified ten dynamically and thermally stable $M$$_{2}$Se, including eight superconductors, four SCTMs, and eight CDW materials, five of which exhibits A-AFM order. All ten stable materials exhibit metallic properties, with the highest superconducting $T_{c}$ reaching 10.4 K, surpassing that of most TMDs and MXenes. The CDW of three non-magnetic materials originates from EPC and can be suppressed by applying appropriate biaxial strain, thereby enabling superconductivity. The $M$$_{2}$Se family combines the elemental composition of TMDs with the stoichiometric ratio of MXenes, successfully preserving and enhancing the diverse properties of these material classes. In summary, our findings introduce a novel family of selenide-based MXene-like 2D materials with various and tunable properties, providing a promising platform for exploring topological superconductivity, CDW phases, and competing quantum orders.

\section{Acknowledgements}
 This work is supported by the National Natural Science Foundation of China (Grant No. 12074213), the National Key R$\&$D Program of China (Grant No. 2022YFA1403103), the Major Basic Program of Natural Science Foundation of Shandong Province (Grant No. ZR2021ZD01), and the Natural Science Foundation of Shandong Province (Grant No. ZR2023MA082).

\bibliography{references}

\end{document}